\documentclass[epj]{svjour}
\usepackage{latexsym}
\usepackage{amsmath,amssymb,amsfonts}
\usepackage[dvips]{graphicx}
\begin{document}
\title{Dynamics of tournaments: the soccer case}
\subtitle{A random walk approach modeling soccer leagues}
\author{H. V. Ribeiro\inst{1,}\thanks{\texttt{hvr@dfi.uem.br}} \and R. S. Mendes\inst{1,2} \and L. C. Malacarne\inst{1} \and S. Picoli Jr.\inst{1} \and P. A. Santoro\inst{1}
}                     
%
%
\institute{Departamento de F\'isica, Universidade Estadual de Maring\'a,  Av. Colombo 5790, 87020-900, Maring\'a, PR, Brazil
\and National Institute of Science and Technology for Complex Systems, CNPq, Brazil}

\date{Received: date / Revised version: date}
%
\abstract{
A random walk-like model is considered to discuss statistical aspects of tournaments.
The model is applied to soccer leagues with emphasis on the scores.
This competitive system was computationally simulated
and the results are compared with empirical data from the English, the German and the Spanish leagues
and showed a good agreement with them. The present approach enabled us to characterize 
a diffusion where the scores are not normally distributed, having a short and asymmetric tail
extending towards more positive values. We argue that this non-Gaussian behavior is related with the
difference between the teams and with the asymmetry of the scores system. In addition, we compared 
two tournament systems: the all-play-all and the elimination tournaments.
} 
\maketitle
\section{Introduction}
In recent years, scientists have become increasingly interested in the behaviors of complex systems
\cite{Auyang,Jensen,Barabasi,Haken}.
Finance\cite{Vandewalle}, genetics\cite{Peng} and religion\cite{Picoli} are just a few examples of areas recently addressed by statistical physicists. However, many of the systems in 
such contexts are not isolated and are sometimes very difficult to describe quantitatively. 
Hence, it is common in these studies to try to capture important features, i.e., universal behaviors.
In this scenery, simple models whose retaining only the main relevant ingredients of the
original systems have been shown to give useful information regarding the underlying processes 
responsible for the observed behavior. For instance, random walk based models have been used to
study aspects from physics and astronomy\cite{Chandrasekhar} to biology\cite{Berg} and economy\cite{Stanley}.

In this work, concepts of random walks are
considered in order to investigate the dynamics of tournaments with emphasis on 
the most popular of them: the soccer tournaments. In this direction,
there is an increasing interest to study the dynamics of soccer.
For instance,  Ref. \cite{Hirotsu} proposes a model to evaluate the 
characteristics of soccer teams such as offensive and defensive strengths. Other works are focused in 
predicting the outcome of sporting contests \cite{Glickman,Ruud,Goddard}. There are also investigations of the 
soccer goal distribution\cite{Dyte,Malacarne,Greenhough,Bittner,Bittner2}, the existence of a home advantage\cite{Clarke},
the persistence in sequences of match results\cite{Dobson}, the temporal sequence of ball 
movements\cite{Mendes2} and the network of Brazilian soccer players\cite{Onody}. As we can see, 
the focus of these works is not to investigate the supposedly universal features of soccer tournaments.
Our present study attempt to fill this hiatus by using a random 
walk-like model which reproduces some of these important features. We basically employ
the statistical framework based on random walk interpretation of soccer leagues
used by Heuer and Rubner\cite{Heuer} (see also Ref. \cite{Ribeiro}). In particular, they argue that the team fitness
is better described by goals difference than number of points. Here, we do not attain to this difference
because we focus only on wins, draws and defeats (i.e. the number of points). Thus, in this degree of
detail our model does not take into account any other aspects such as match scores or dynamic of
ball movements that have been related with Poissonian processes\cite{Ruud,Mendes2}.

The organization of this paper is as follows. In Section 2 we present our observational data and 
some statistical analysis. In Section 3 we present our model for the scores and compare it 
with the observational data. We also explore the model to characterize the soccer leagues. In Section 4
we look for other observational quantities: the number of wins, draws and defeats. In section 5, by using
the model, we compare two kinds of tournament system: the all-play-all system and the elimination tournaments.
Finally, in Section 6, we present a summary and some concluding comments.

\section{Data analysis}

\begin{figure*}[t!]
   \begin{center}
	\includegraphics[scale=0.55]{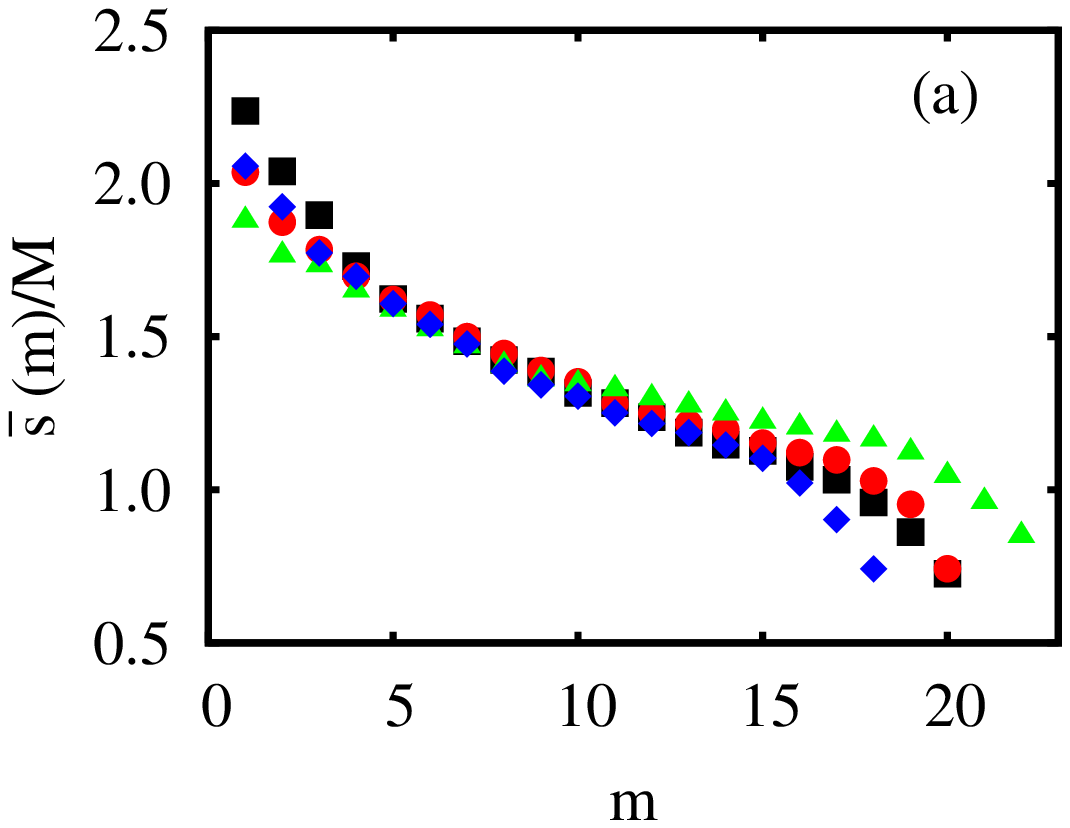}
	\includegraphics[scale=0.55]{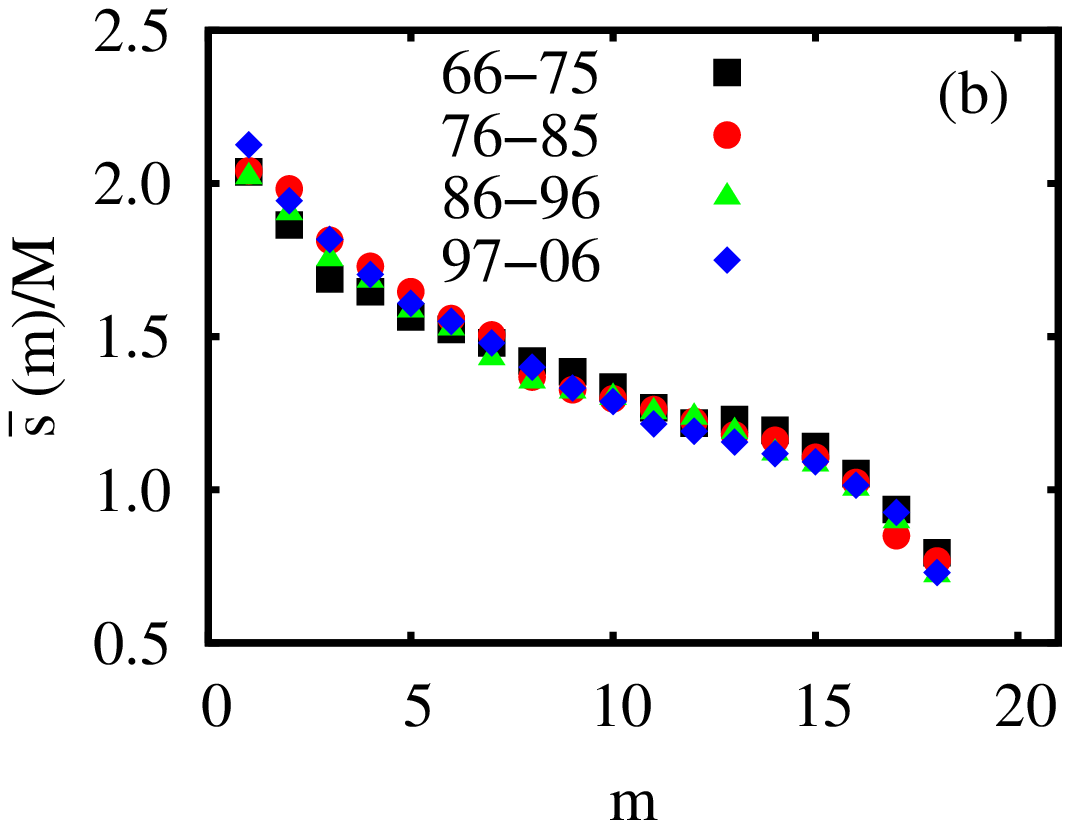}
	\includegraphics[scale=0.55]{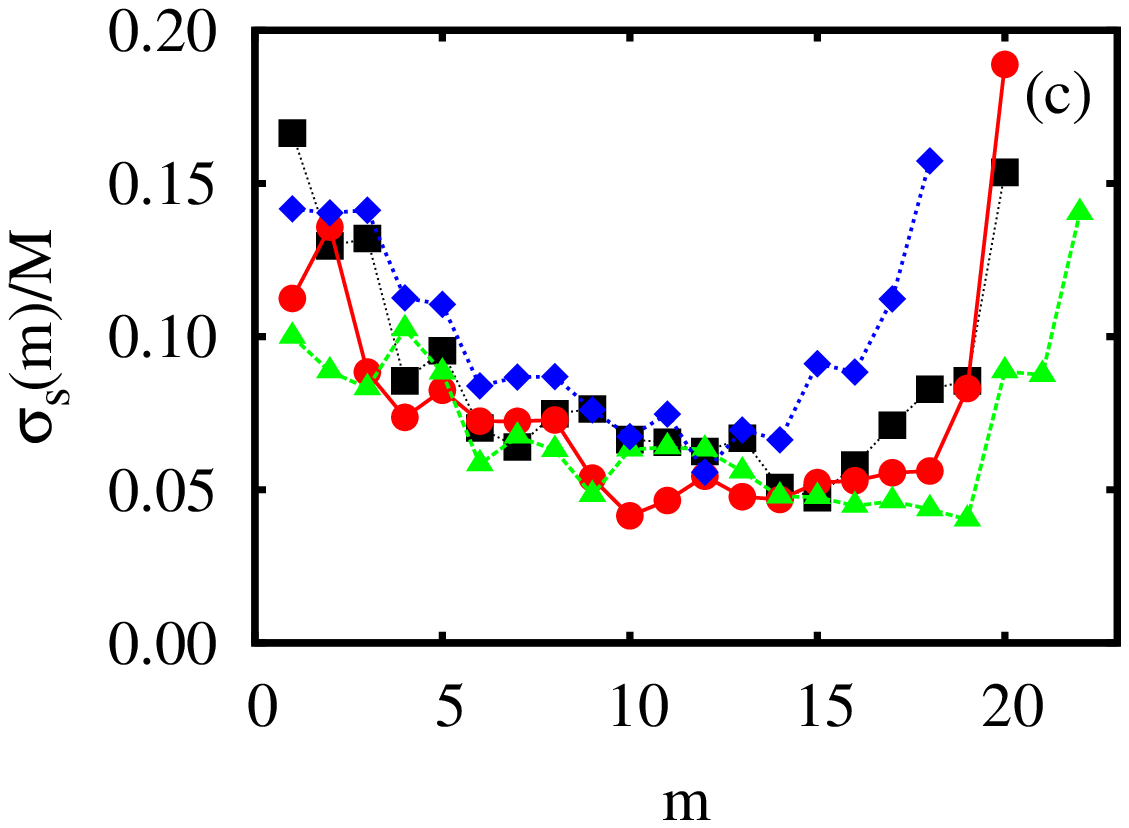}
	\caption{(a) $\bar{s}(m)$ and (c) $\sigma_s(m)$ for 
		German Bundesliga (diamonds), English League (squares), Spanish League A (circles)
		and Spanish League A (triangles). (b) $\bar{s}(m)$ for several periods of German Bundesliga.}
	\label{fig:data}
   \end{center}
\end{figure*}

As database we have taken the results from the German Bundesliga\cite{FedAlema} in the period from 1965 to 2007
(41 seasons, except the season 1991/92 because it contained more than 18 teams), from the English 
League\cite{FedIngles} (12 seasons, 1995-2007), from the Spanish League A\cite{FedEspanha} (11 seasons, 1996-2007)
and from the Spanish League B\cite{FedEspanha} (9 seasons, 1998-2007).

In every tournament every team plays against all the others twice in each season, once at home and once away,
totaling $M$ matches.
The standard score system used in many sports leagues, especially in soccer tournaments, 
states 3 points for a win, one point for a draw and no points for a defeat. However, the score
system we found was different: in the past it stated 2 points for a win, rather than 3. The year when
the ``3 points for a win'' was adopted is different for each league. In our data set, only the 
results from the German Bundesliga mix the two score systems. This league adopted the current system in 1995,
the year when FIFA (F\'ed\'eration Internationale de Football Association) formally adopted the new system, which became standard in international tournaments,
as well as most national soccer leagues. Hence, to ensure that our data have a unique score system,
we recalculate the scores from German Bundesliga using the actual scheme.

We will represent the score of the rank $m$ in the season $i$ in the round number $r$ as $s_i(m,r)$.
This quantity is like a microscopic measurement, and therefore is subject to fluctuations. 
In order to minimize the fluctuations, 
we start investigating the average of this quantity over the seasons in the final round $r_{f}$, i.e.,
\begin{equation}
  \bar{s}(m) = \frac{1}{N} \sum_{i=1}^{N} s_i(m,r_{f}),
\end{equation}
where $N$ is the number of seasons.
Figure \ref{fig:data}a shows $\bar{s}(m)$ for all the empirical data set. We can see that $\bar{s}(m)$ presents
a similar shape for all leagues, even though the number of teams is different. Someone could argue about
a possible time-dependent behavior of this shape. In order to verify it we calculated $\bar{s}(m)$
over four distinct periods of Germany Bundesliga and, as shown in Figure \ref{fig:data}b, the shape 
does not change significantly.

Next, we investigate the standard-deviation of the variable $s_i(m,r_{f})$ as a rank function:
\begin{equation}
 \sigma^2_s(m)=\frac{1}{N-1} \sum_{i=1}^{N} [s_i(m,r_{f})-\bar{s}(m)]^2.
\end{equation}
Figure \ref{fig:data}c shows this quantity for all empirical data. Again, we found a similar shape for all leagues.
Note that these fluctuations are larger for both extremal ranks. A similar result has been recently reported
for the average win fraction of baseball\cite{Sire}.
\begin{figure*}[ht!]
   \begin{center}
	\includegraphics[scale=0.55]{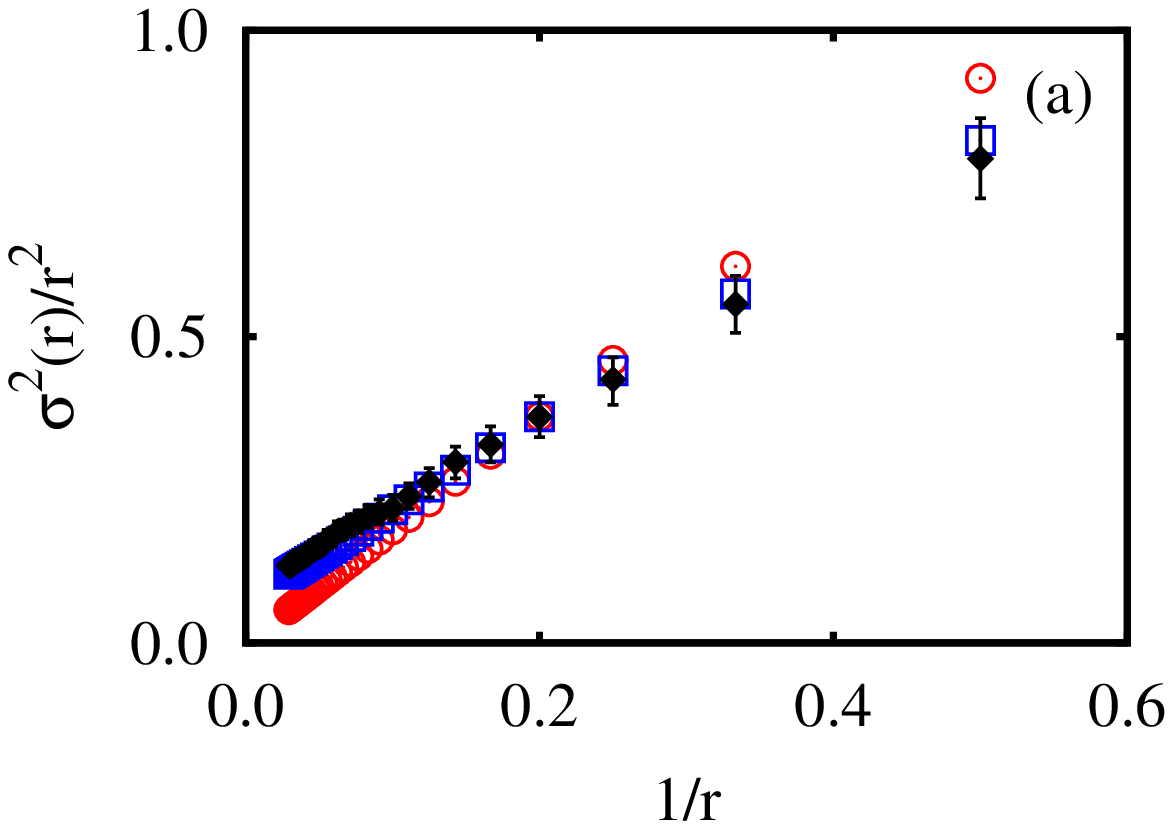}
	\includegraphics[scale=0.55]{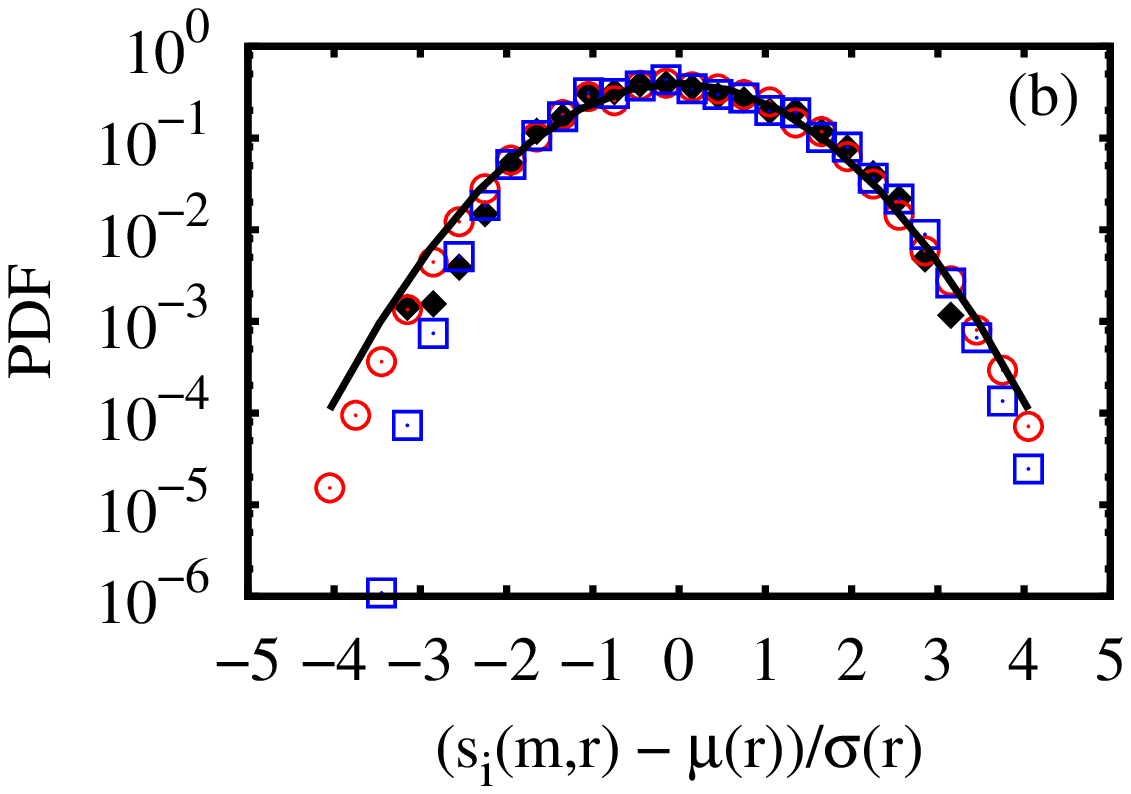}
	\includegraphics[scale=0.55]{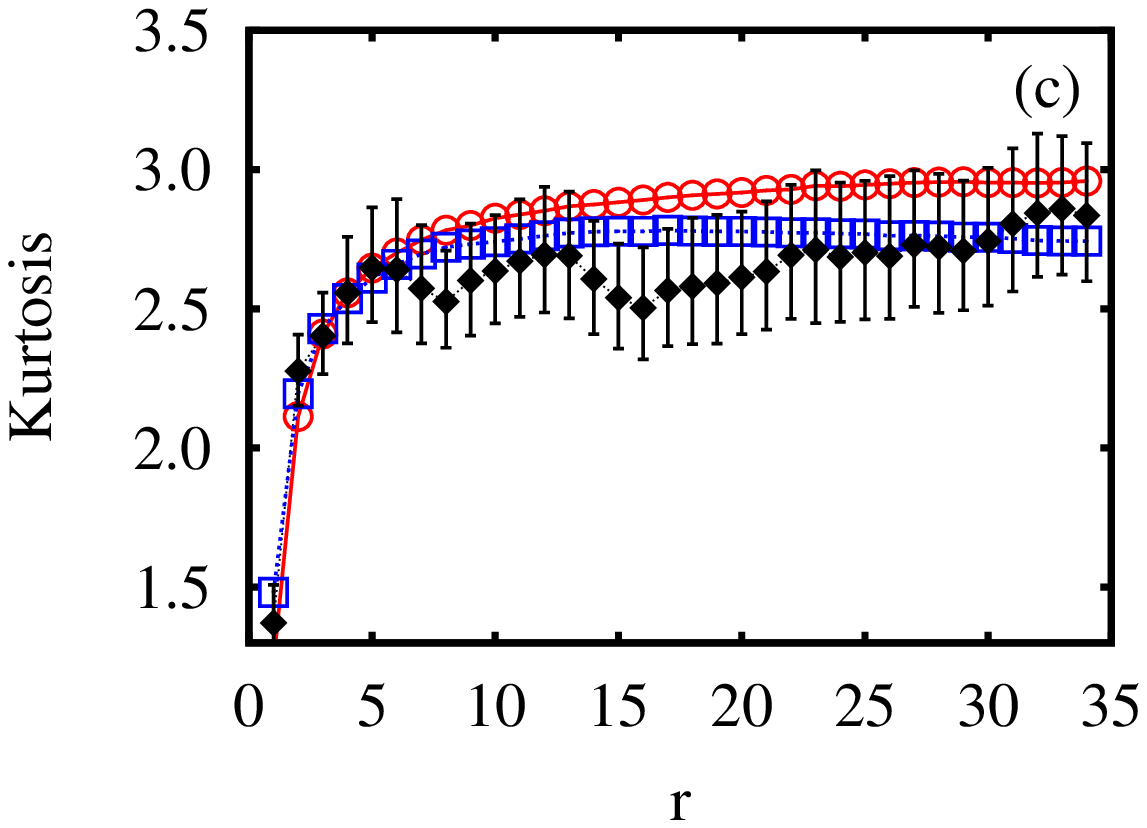}
	\includegraphics[scale=0.55]{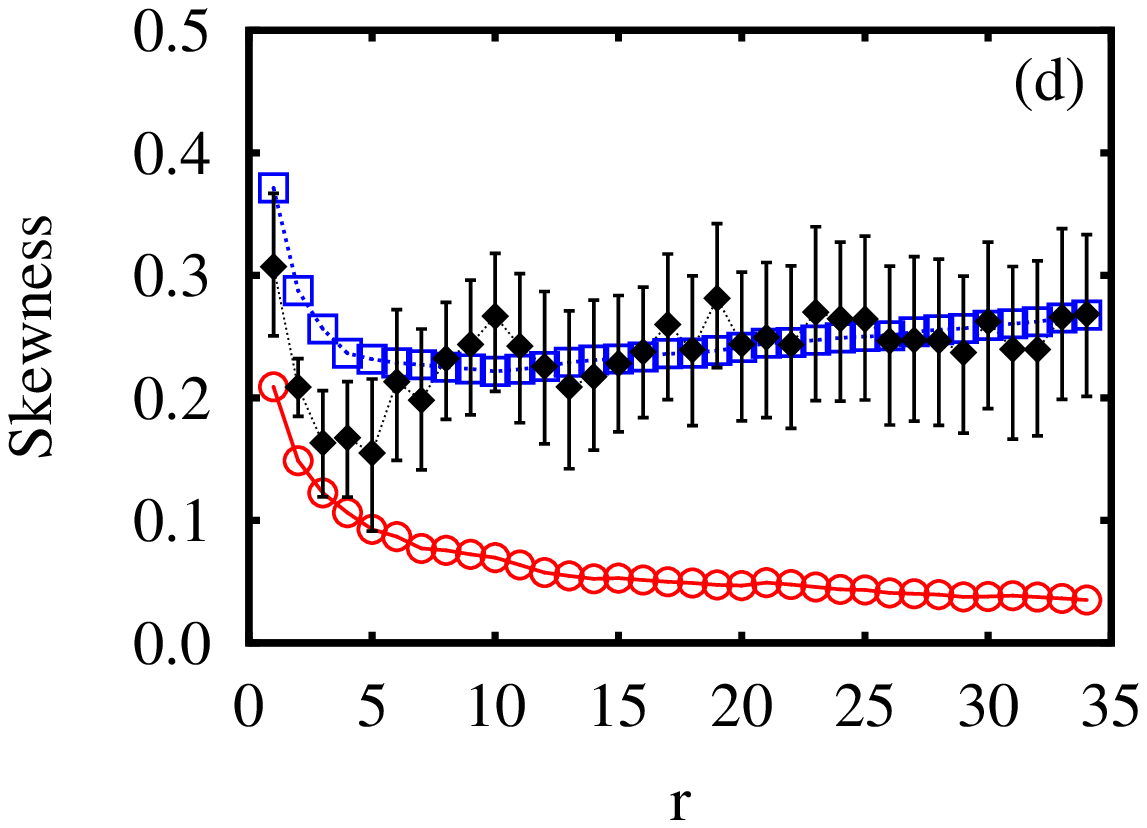}
	\caption{Statistical analysis of Germany Bundesliga; observational data (diamonds), minimalist model (circles) and 
non-identical team model (squares). (a) Variance $\sigma^2(r)$. 
(b) Probability distribution of scores (PDF) compared with a Gaussian (continuous line). The coefficient of (c) kurtosis and (d) skewness versus $r$. The error bars are calculated via bootstrap resampling method \cite{Efron}.}
	\label{fig:data_stats}
   \end{center}
\end{figure*}

This analysis can be extended to a microscopic point of view if we consider the soccer or any other
leagues as set of erratic trajectories. A possible manner to establish this correspondence 
is supposing that the teams are like particles and the scores are the positions.
The ``motion'' is governed by the match result. At each round (time) a team (particle) can jump
three units of length to the right (if the team wins) or one unit of length to the right (if the team draws)
or stay in the same position (if the team loses). We have, therefore, a random walk-like process: 
for each team, at the first round, there are only 3 allowed ``positions'' ($0,1,3$), at the second 
round there are 6 ($0,1,2,3,4,6$), and so on.

This type of analysis needs a larger number of data. Thus, we will use the results from Germany Bundesliga,
because it has more data. We start investigating the standard-deviation over all seasons as a function
of the round number $r$, given by
\begin{equation}\label{eq:std}
 \sigma^2(r)=\frac{1}{N T-1} \sum_{i=1}^N \sum_{m=1}^T (s_i(m,r)- \mu(r) )^2\,,
\end{equation}
where 
\begin{equation}
\mu(r) = \frac{1}{N T} \sum_{i=1}^N \sum_{m=1}^T s_i(m,r)\nonumber
\end{equation}
and $T$ is the number of teams. Figure \ref{fig:data_stats}a shows $\sigma^2/r^2$ versus $1/r$.
In this representation, the diffusive process can be interpreted as usual random walk with a drift, i.e.,
\begin{equation}\label{eq:heuer}
  \frac{\sigma^2(r)}{r^2} =  a + \frac{b}{r}
\end{equation}
where $a$ and $b$ are essentially variances related with the teams' fitness
and statistical fluctuations\cite{Heuer}. Here we have $a \approx 0.075$ and $b \approx 1.55$.


We also evaluate the probability distribution function (PDF) of the scores 
(for the collapsed data, i.e., all seasons of the Germany Bundesliga with all rounds together) and compare it
with a Gaussian distribution as shown in Figure \ref{fig:data_stats}b. This result indicates that the scores are not normally distributed, having a short and asymmetric tail extending towards more positive values. This feature becomes more evident when we look for the kurtosis (Figure \ref{fig:data_stats}c) and skewness (Figure \ref{fig:data_stats}d) coefficients\cite{Statistics}. The asymmetric tail reflects the asymmetry in the score system, i.e.,
the 3 points for a win. We will see that (from our simulations results) if the winner sums 2 points rather than 3 
the skewness tends to zero.

\section{Modelling}

In the previous session, we presented some observational features of the soccer leagues. 
Now, we will firstly present a minimalist model and subsequently another one which 
reproduces these observational behaviors very well.

\subsection{A minimalist model}
As a first attempt to model the previous results, we can use a mean-field-like approximation. In this approach, we
consider that all teams are identical and that the match results are obtained from
a simulation algorithm. The procedure to simulate a match between two teams $i$ and $j$
starts drawing two uniform random numbers, $x_i$ and $x_j$, in the interval [0,1].
Thus, we use them in the following algorithm:
\begin{eqnarray}\label{algo}
\textit{\begin{tabular}{l}
IF $|x_i-x_j|\leq\delta$  \\ 
\hspace{0.5cm}the game ends in a draw;  \\ 
ELSE \\
\hspace{0.5cm}IF $x_i>x_j$ \\
\hspace{0.5cm}\hspace{0.5cm}the winner is the team $i$;\\
\hspace{0.5cm}ELSE \\
\hspace{0.5cm}\hspace{0.5cm}the winner is the team $j$;\\
\end{tabular}}
\end{eqnarray}
of which the outcome of the game emerges. If we were considered other tournaments
(for instance, basketball) where there are no draws, the first step of this algorithm would have to be eliminated.

Employing this procedure, we simulated an entire season $10^5$ times.
This minimalist model has only one parameter $\delta$ associated with the draws. We incrementally
update the values of $\delta$ to minimize, via the method of least squares, the difference between the simulated values of $\bar{s}(m)$
and the observational ones. The best values for this parameter are close together: $\delta=0.11$ for
German Bundesliga and the Spanish League A; $\delta=0.15$ for the Spanish League B and $\delta=0.10$
for the English League.

\begin{figure*}[ht!]
   \begin{center}
	\includegraphics[scale=0.55]{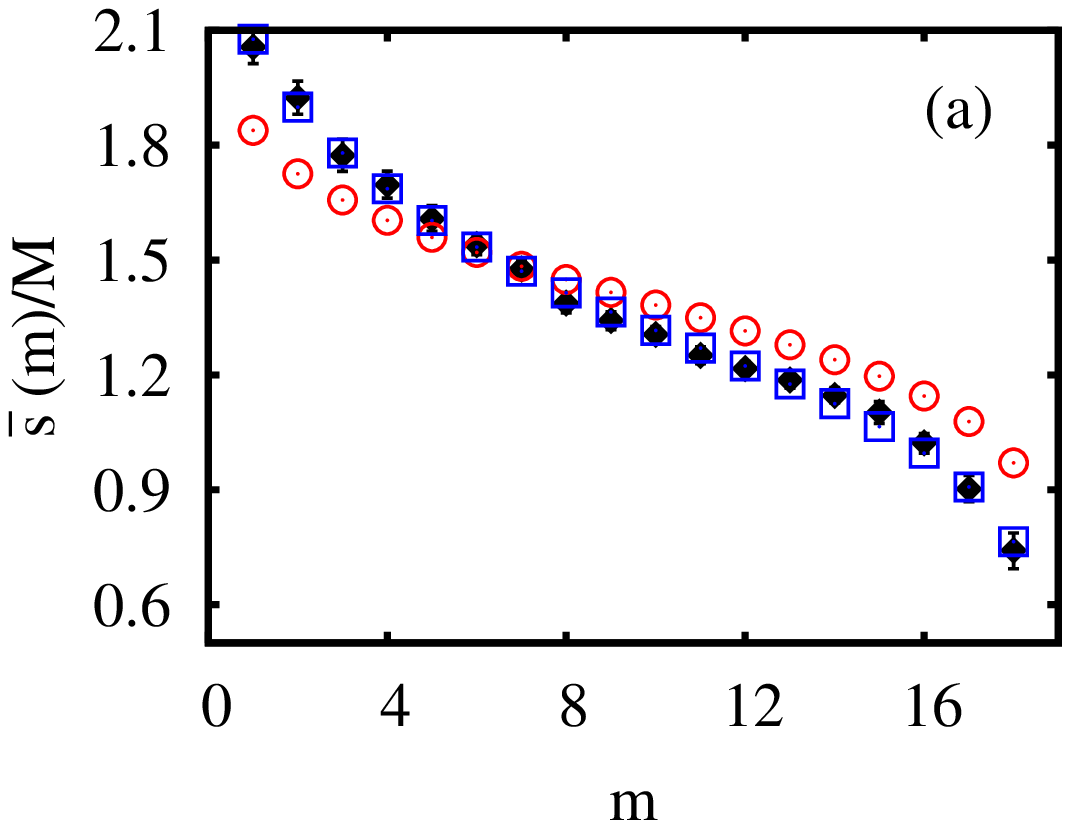}
	\includegraphics[scale=0.55]{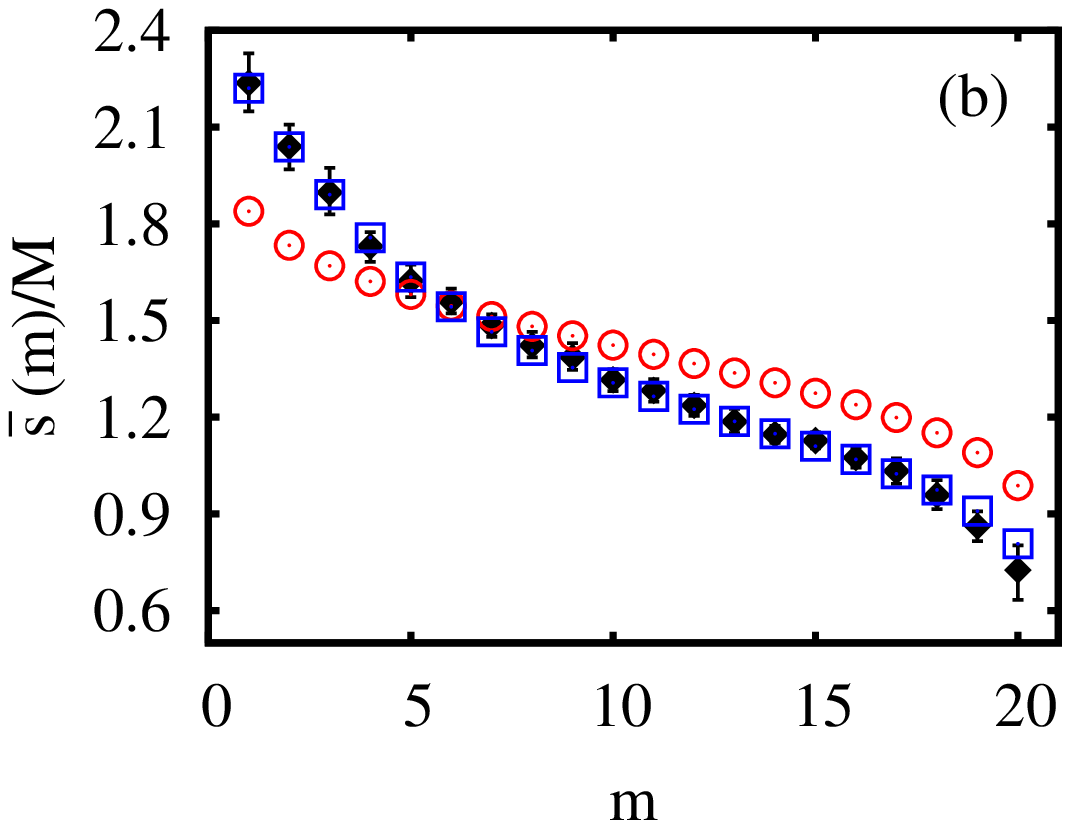}
	\includegraphics[scale=0.55]{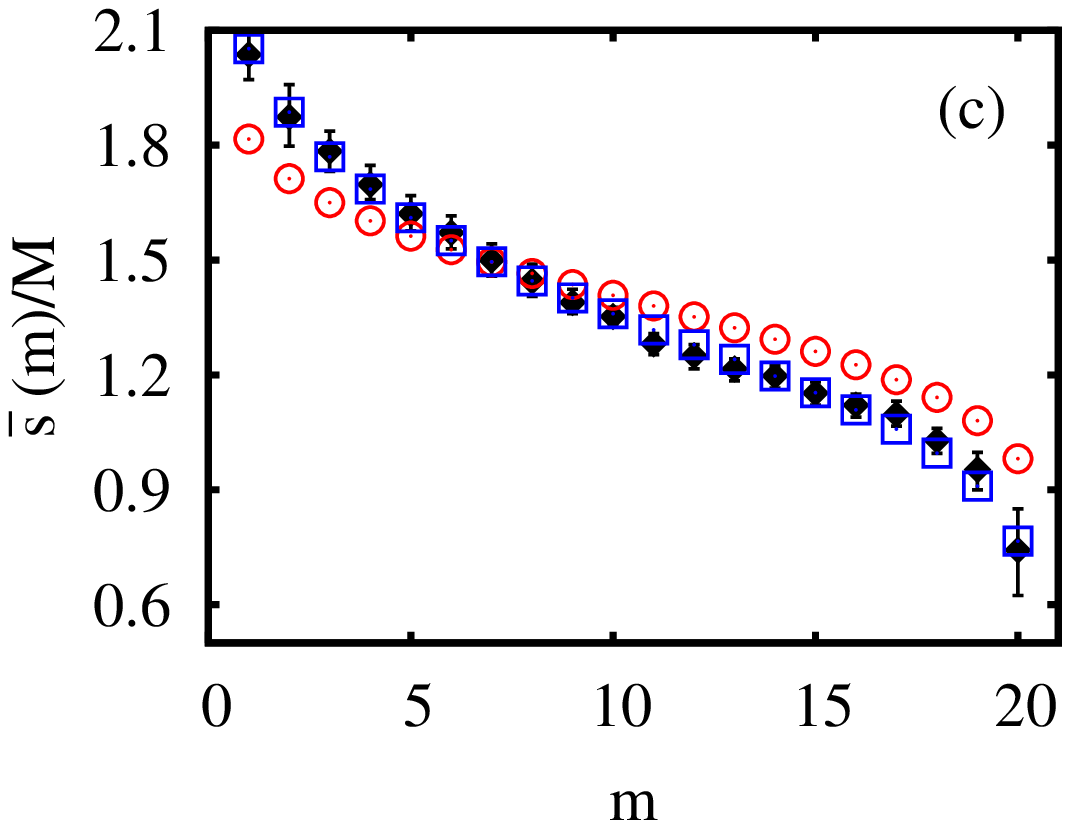}
	\includegraphics[scale=0.55]{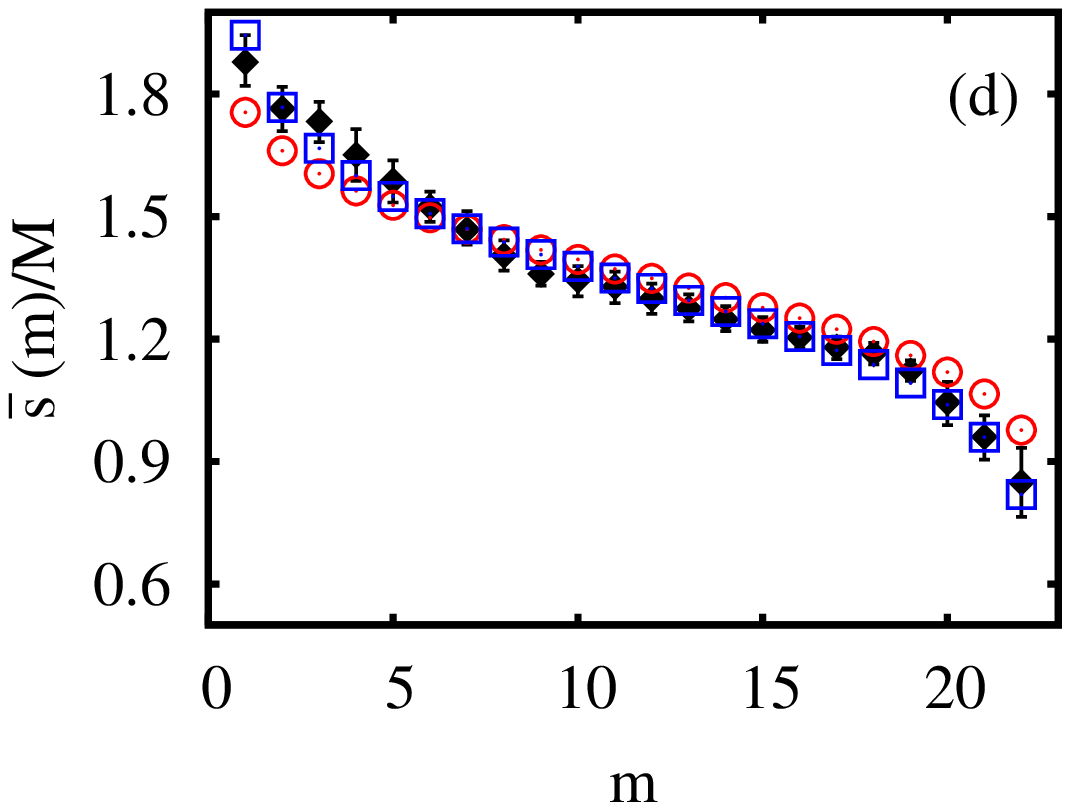}
	\caption{Average rank score: a comparison between observational data (diamonds), minimalist model (circles) and non-identical team model (squares) for (a) German Bundesliga, (b) English League, (c) Spanish League A and (d) Spanish League B.}
	\label{fig:fit}
   \end{center}
\end{figure*}

\begin{figure*}[ht!]
   \begin{center}
	\includegraphics[scale=0.55]{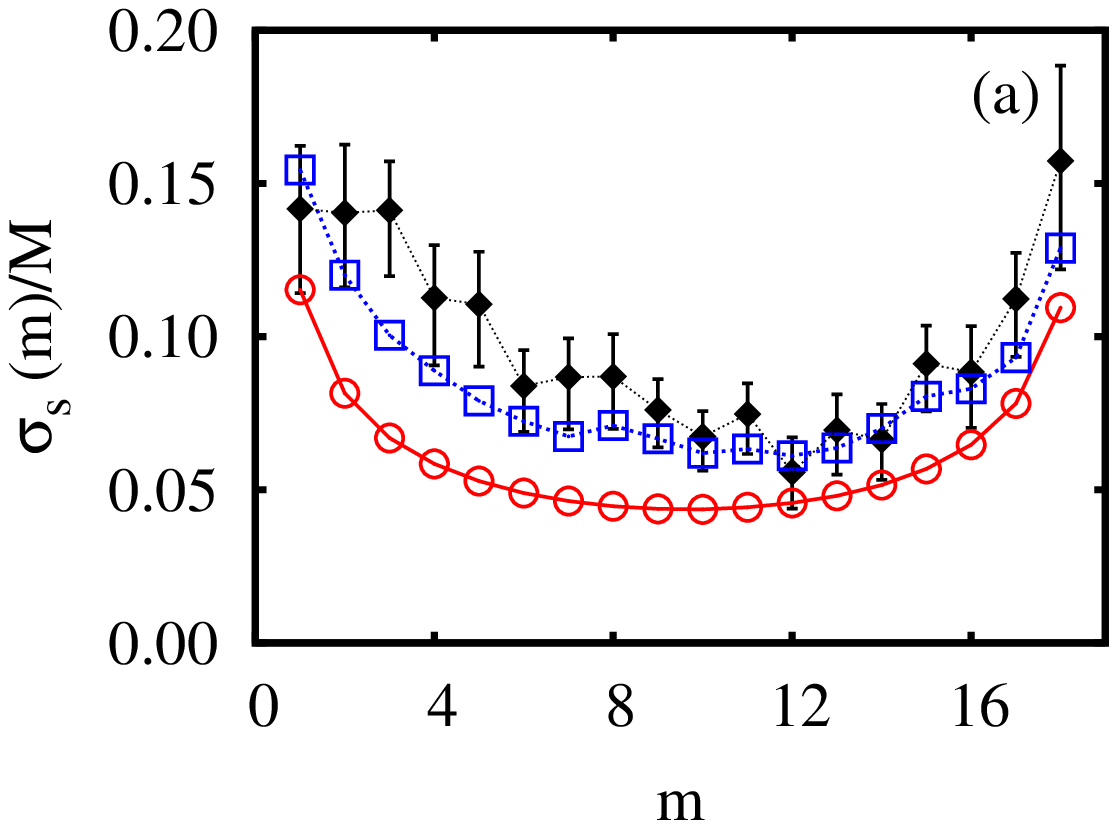}
	\includegraphics[scale=0.55]{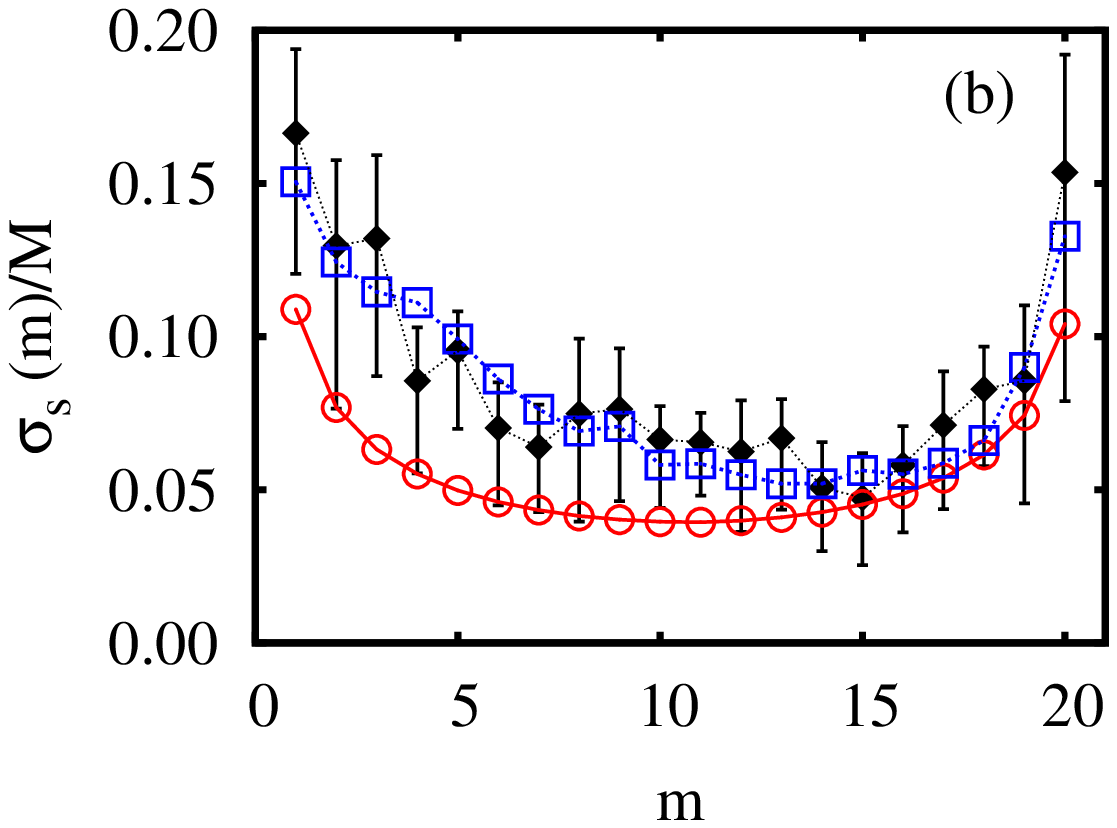}
	\includegraphics[scale=0.55]{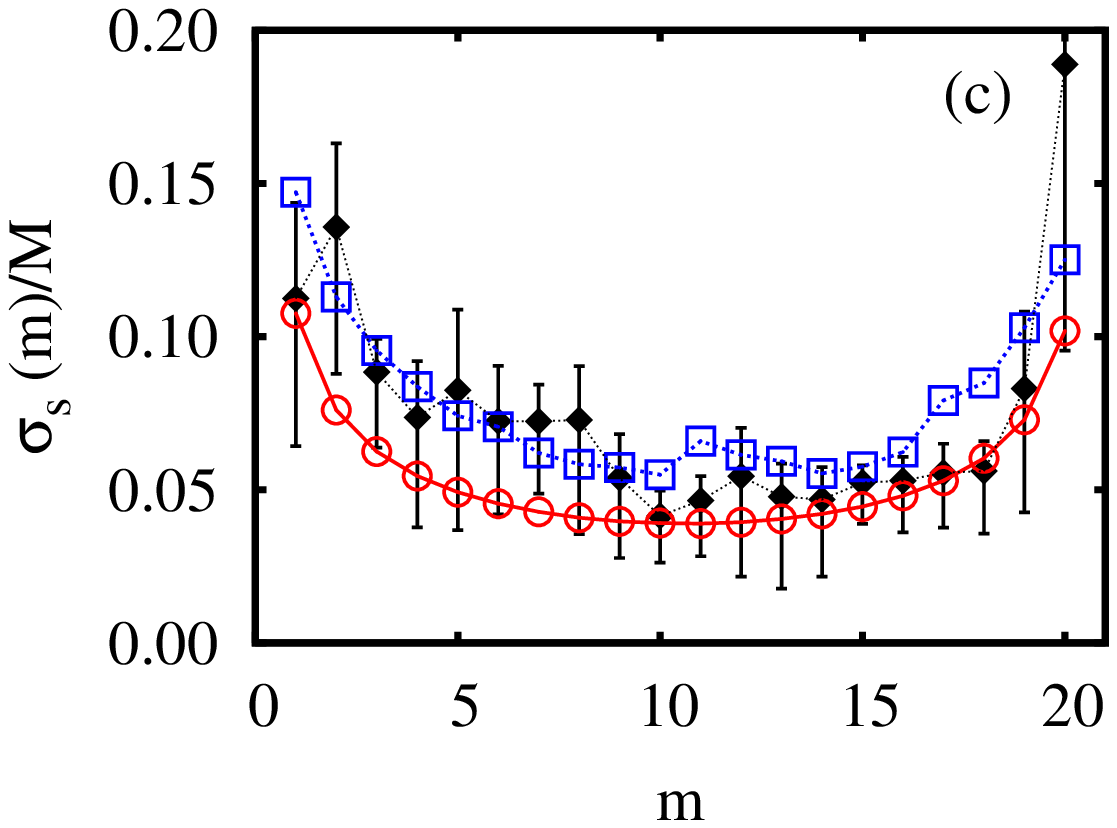}
	\includegraphics[scale=0.55]{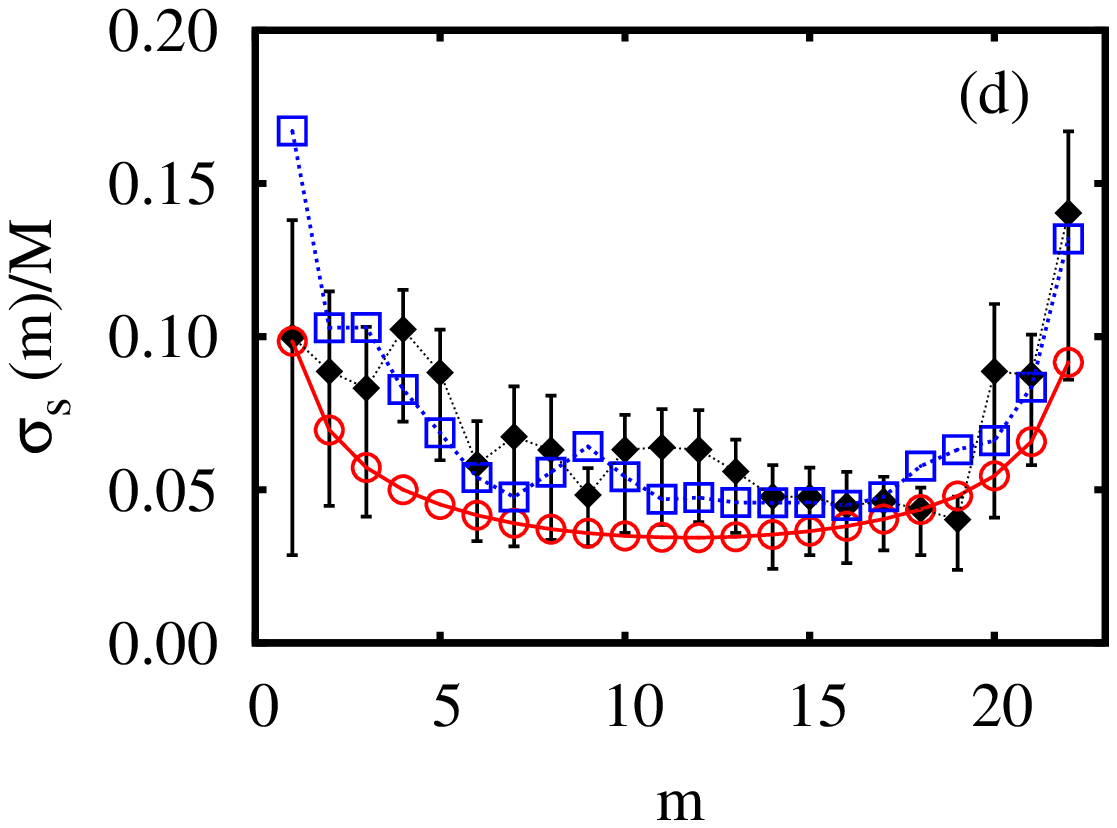}
	\caption{Standard-deviation of the rank scores: a comparison between observational data (diamonds), minimalist model (circles) and non-identical team model (squares) for (a) German Bundesliga, (b) English  League, (c) Spanish League A and (d) Spanish League B.}
	\label{fig:data_sd}
   \end{center}
\end{figure*}

A comparison between simulated and observational data is shown in Figure \ref{fig:fit} and, as we can see, this minimalist
model can not explain the behavior of $\bar{s}(m)$ found in the empirical data. The discrepancy
between model and simulation is larger in the English League and smaller in the Spanish League B. We
will see that these discrepancies are related with the difference between the teams, which are greater in the English
League than in the Spanish League B. We also evaluate $\sigma^2_s(m)$ from this model and, as shown in Figure \ref{fig:data_sd}, it only describes the data behaviour qualitatively.

For completeness, we also evaluate the standard-devia\-tion $\sigma(r)$, the PDF and the coefficients
of kurtosis and skewness shown in Figure \ref{fig:data_stats}. Since this model is a mean-field-like approximation,
we do not expect it to reproduce the non-Gaussian behavior present in the data. In fact, for this case,
the central limit theorem
states that the PDF of the sum of many independent random variables tends to a normal distribution.
This fact becomes evident if we note that kurtosis and the skewness tend to the normal values
(see Figures \ref{fig:data_stats}c and \ref{fig:data_stats}d).

\subsection{A non-identical team model}
\begin{figure*}[ht!]
   \begin{center}
	\includegraphics[scale=0.55]{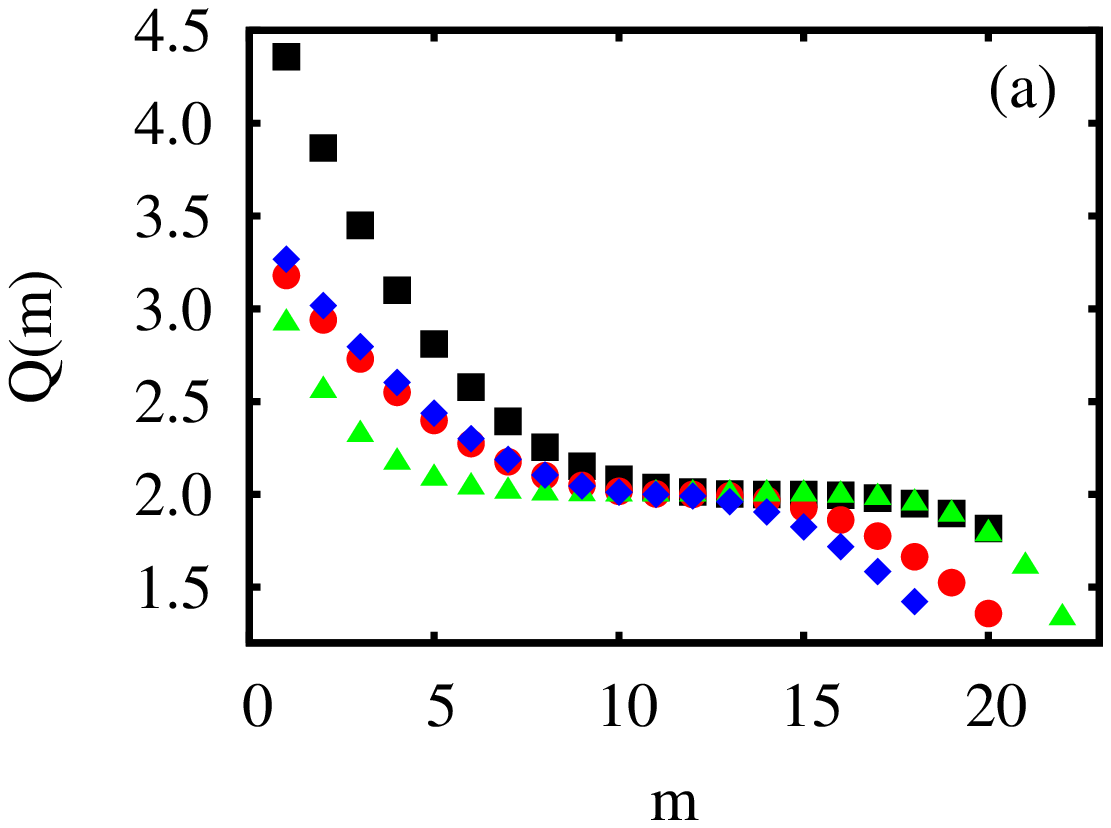}
    \includegraphics[scale=0.55]{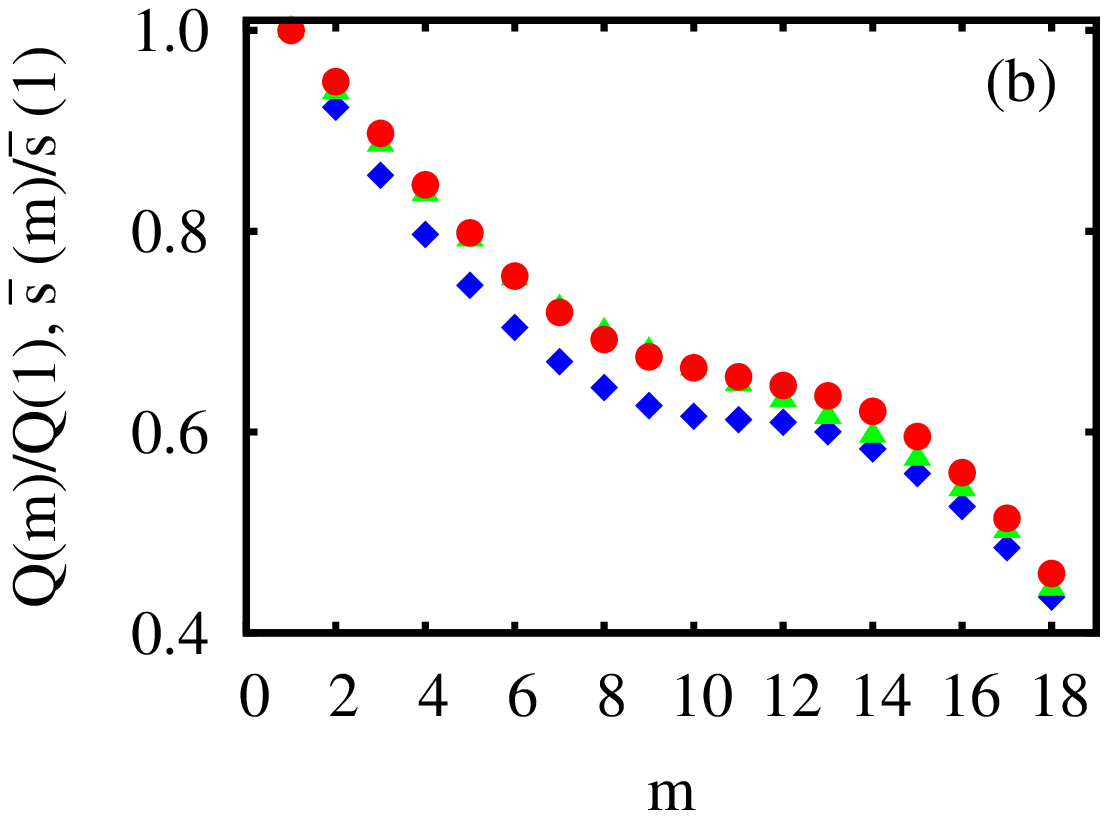}
	\caption{(a) The quality factors $Q(m)$ that emerge from non-identical team model for 
		German Bundesliga (diamonds), English League (squares), Spanish League A (circles)
		and Spanish League B (triangles). (b) A comparative presentation of $Q(m)$ (diamonds)
		and rank average for 10 rounds (triangles) and 100 rounds (circles).}
	\label{fig:quality}
   \end{center}
\end{figure*}
Now, we need some additional ingredients to describe the discrepancies 
previously indicated. Evidently, in a soccer match, the final result is governed by many unforeseen
effects which are generally very difficult to model. Thus, we effectively focus our attention on the difference between the teams.
It is common sense that the teams are non-identical;
for example, they have different offensive and defensive strengths. But how to model these differences? 
For instance, one soccer report could suggest that the teams are divided into two groups: 
the small and the big teams.  

In this work, as a first approximation, we suppose that each team is fully characterized
by only one parameter $Q$, i.e., a single quality factor. 
Recently, by considering the Bundesliga, a careful investigation pointed that the quality of
a team is better described by goal difference than the number of points and that it is
constant over each season\cite{Heuer}. However, due to minimalist approach desired in this work,
our model does not take some particular ingredients into account, such as,
temporal variations over the seasons, home advantage, team-specific characterization, and 
specific details about the quality factor ({\it e.g.}, goals difference instead of number of points).
Moreover, we emphasize that $Q$ refers to an average behavior of several championships,
therefore $Q$ is not related with a specific team nor with its rank within a particular tournament.

It is desirable to employ a functional form of $Q$ with few parameters 
in contrast with the many parameters (the number of teams) necessary to fully specify $Q$. 
In this direction, we note that our data (see Figures \ref{fig:data}a and \ref{fig:data}c)
remember the shape of $f(x)\sim -|x|^\alpha$ shifted. Thus, in order to take 
these aspects into account and to overcome possible divergences, 
our guess is to assume that $Q$ versus rank ($m$) has the very adjustable functional form
\begin{eqnarray}\label{eq_Q(m)}
 Q(m) &=& 2 + \frac{T -2 m - \epsilon - \beta/2}{T} \nonumber \\ &\times&
  \left \vert \left( {\frac{T - 2 m - \epsilon - \beta/2}{T}}\right) ^{\alpha -1}\right\vert\,,
\end{eqnarray}
where $\epsilon$ is a very small number ($\epsilon\ll1$), $\alpha \geq 0$ and $\beta$ are parameters that dictate the $Q(m)$ form. When $\alpha$ is zero the teams are distributed into two groups ($Q_1=1$ and $Q_2=3$). 
The increasing of $\alpha$ begins to distinguish
the teams in a continuum. The curve translates to the right if $\beta<0$ or to the left if $\beta>0$
and, when $\beta=0$, the function $Q(m)-2$ is odd with respect to $m=T/2$.

In the direction of a simple model, we would like to emphasize that the choice of $Q(m)$ 
via eq. (\ref{eq_Q(m)}) is an attempt,
motivated by the data, to promote a good adjustment of the model by using a minimal number
of parameters. Of course, other forms for $Q(m)$ may be employed, for instance, we could 
also consider it based on a normal distribution, log-normal distribution\cite{James} or
linear combination of two normal distributions\cite{Heuer}. When considering our data, 
these possibilities do not give a significant improvement in the results.

In order to use this quality factor, we make a change in the previous simulation algorithm (\ref{algo}).
The two random numbers $x_i$ and $x_j$ are now respectively distributed in
the interval $[0,Q(i)]$ and $[0,Q(j)]$, where $Q(i)$ is the quality factor of the team $i$ and $Q(j)$ is the
same for the team $j$, in addition, to relativize the parameter $\delta$ we replaced it for 
$\delta'=\delta \frac{Q(i)}{Q(j)}$ with $i>j$. Thus, we are working with the constrained random walk
\begin{equation}
s(i,r+1)=s(i,r)+\xi_{i}\,,
\end{equation}
where $\xi_{i}=3$ (1 or 0) if the team $i$ wins (draws or defeats) at $(r+1)$th round
and for the team $j$ we have $\xi_{j}=0$ (1 or 3) due to the constraint.

We performed the simulation varying the model parameters ($\alpha$, $\beta$ and $\delta$) to minimize,
via the method of least squares, the difference between the simulated values of $\bar{s}(m)$ and the data set ones. 
The best values for the
parameters are shown in Table \ref{tab:parameters}. Note that, besides the changes of statistical
properties of German Bundesliga during more than 40 years, the parameters $\alpha$ and $\delta$ corresponding to the
last decade (1997-2007) are similar. This model not only reproduces $\bar{s}(m)$ very well
(see Figure \ref{fig:fit}), but it also correctly describes the behavior of the standard-deviation $\sigma^2_s$ 
(see Figure \ref{fig:data_sd}).



\begin{table}
\centering
\caption{The best values for the parameters.}
\label{tab:parameters}
\begin{tabular}{llrrr}
\hline\noalign{\smallskip}
League  & Period &  $\alpha$ & $\beta$ & $\delta$\\
\noalign{\smallskip}\hline\noalign{\smallskip}
German Bundesliga & 1965-2007 & 2,10 &  -8,29 & 0,40\\
German Bundesliga & 1997-2007 & 2,08 &  -12,00 & 0,42\\
English League    & 1995-2007 & 2,99 & -17,29 & 0,45\\
Spanish League A  & 1996-2007 & 2,33 &  -6,94 & 0,41\\
Spanish League B  & 1998-2007 & 5,18 &  -3,30 & 0,40\\
\noalign{\smallskip}\hline
\end{tabular}
\end{table}

Further aspects of the random walk-like process described in Section 2 are in very good
agreement with the present model. In fact, as shown in Figure \ref{fig:data_stats}, the present model 
explains the behavior of the variance $\sigma^2(r)$, kurtosis, skewness and the PDF.

From our fit, we have obtained the functional form of $Q(m)$ shown in Figure \ref{fig:quality}a. This function 
gives us some
information about the championship competitiveness. In the previous section, we saw that the minimalist model
gives a better agreement for the Spanish League B data. Now, we can note that the shape of $Q(m)$ for
this case has more teams in the same baseline than all others. This result indicates that the Spanish League B
is the most balanced league from all the empirical data set. On the other hand, the English League presents the most
different shape. Note that the quality factors of the first ranks are substantially greater than the others.
Therefore, the minimalist model gives poor agreement for this league. This result suggests that in this league
there are some teams which are very strong. In fact, from 1995 to 2007, only three teams won this championship\footnote{Manchester United, Arsenal and Chelsea.}, unlike there are eleven different champions in the same period of the Spanish League B\footnote{Las Palmas, Sporting B, Cacere\~no, Getafe, Universidad LPGC, Atl\'etico B, Barakaldo, Universidad LPGC, Pontevedra, Real Madrid B and Pontevedra.}.
\begin{figure*}[ht!]
   \begin{center}
	\includegraphics[scale=0.55]{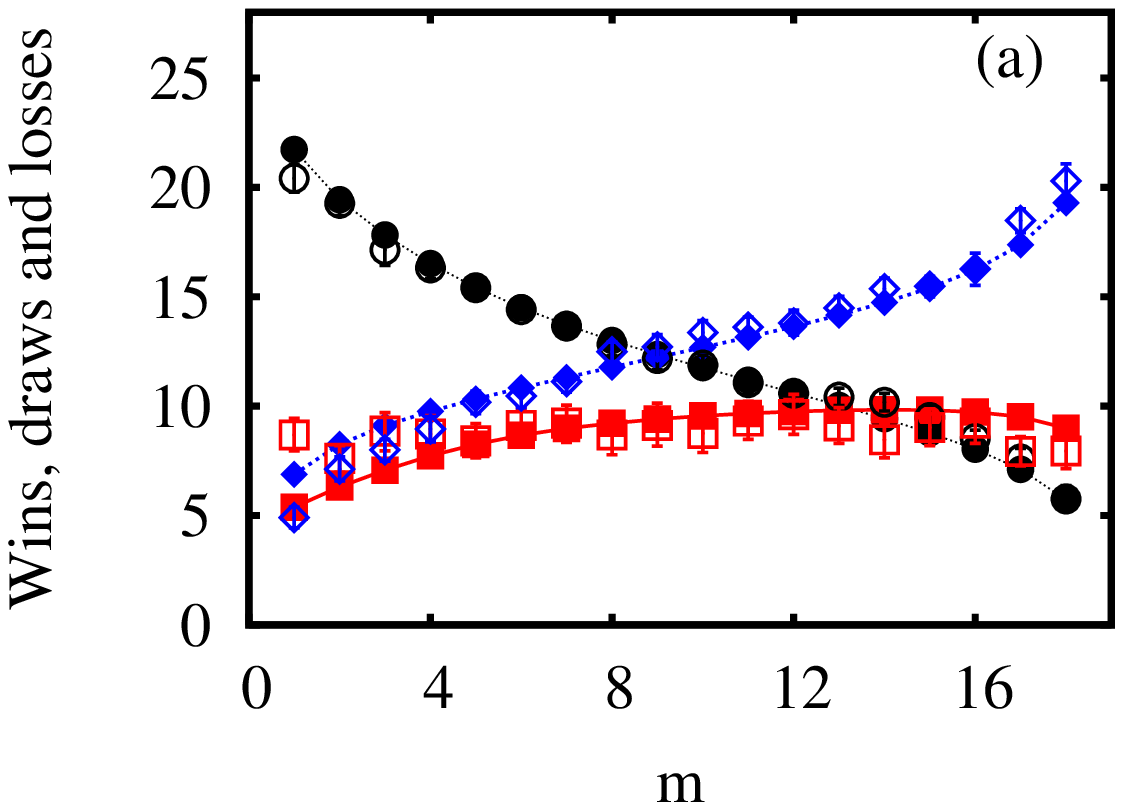}
	\includegraphics[scale=0.55]{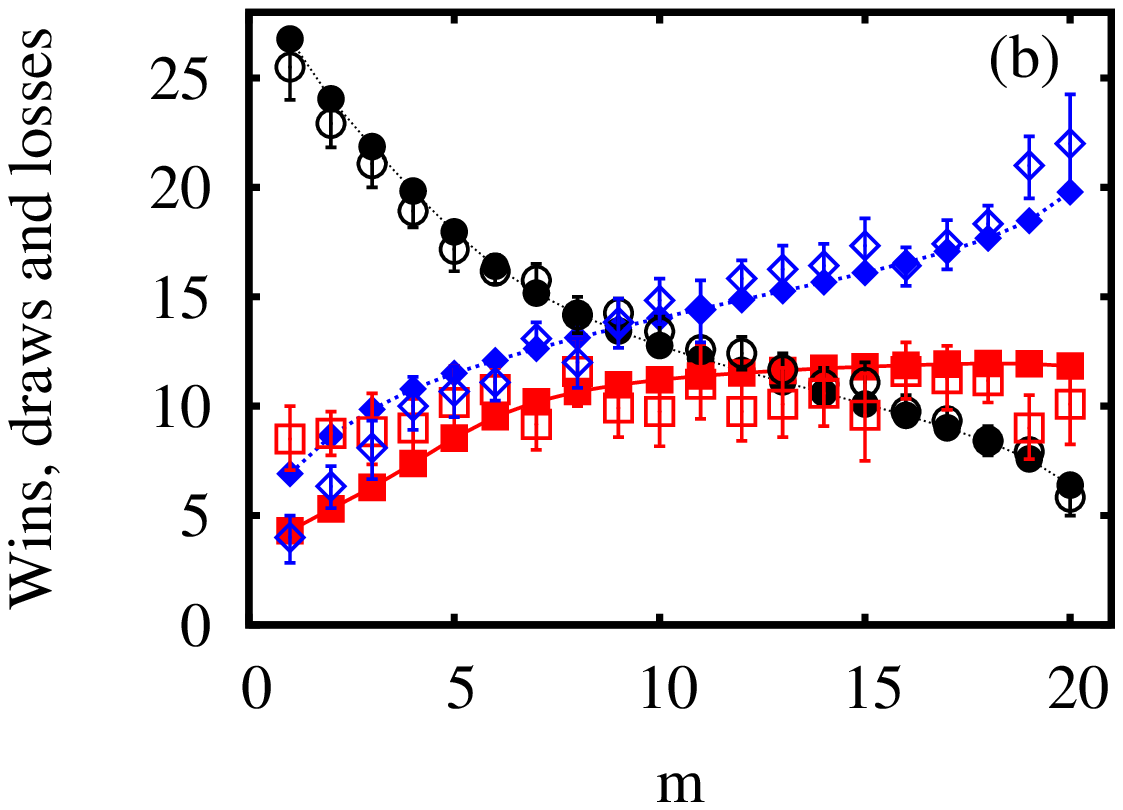}
	\includegraphics[scale=0.55]{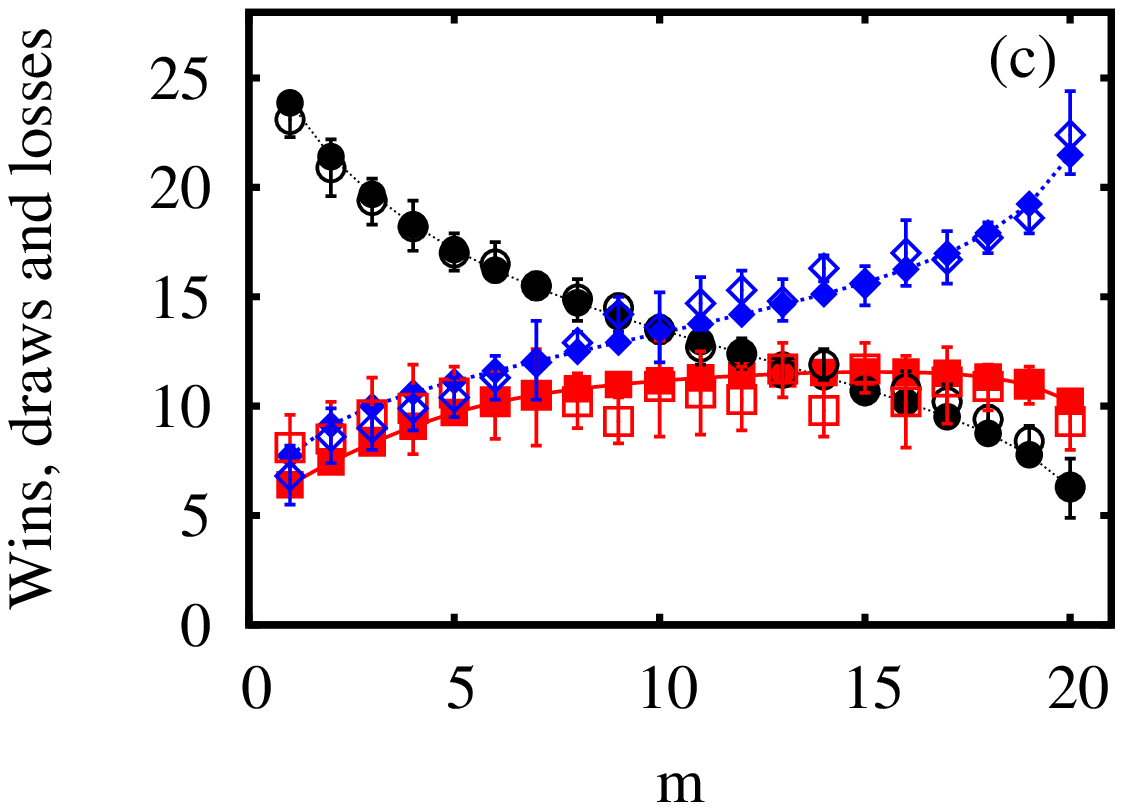}
	\includegraphics[scale=0.55]{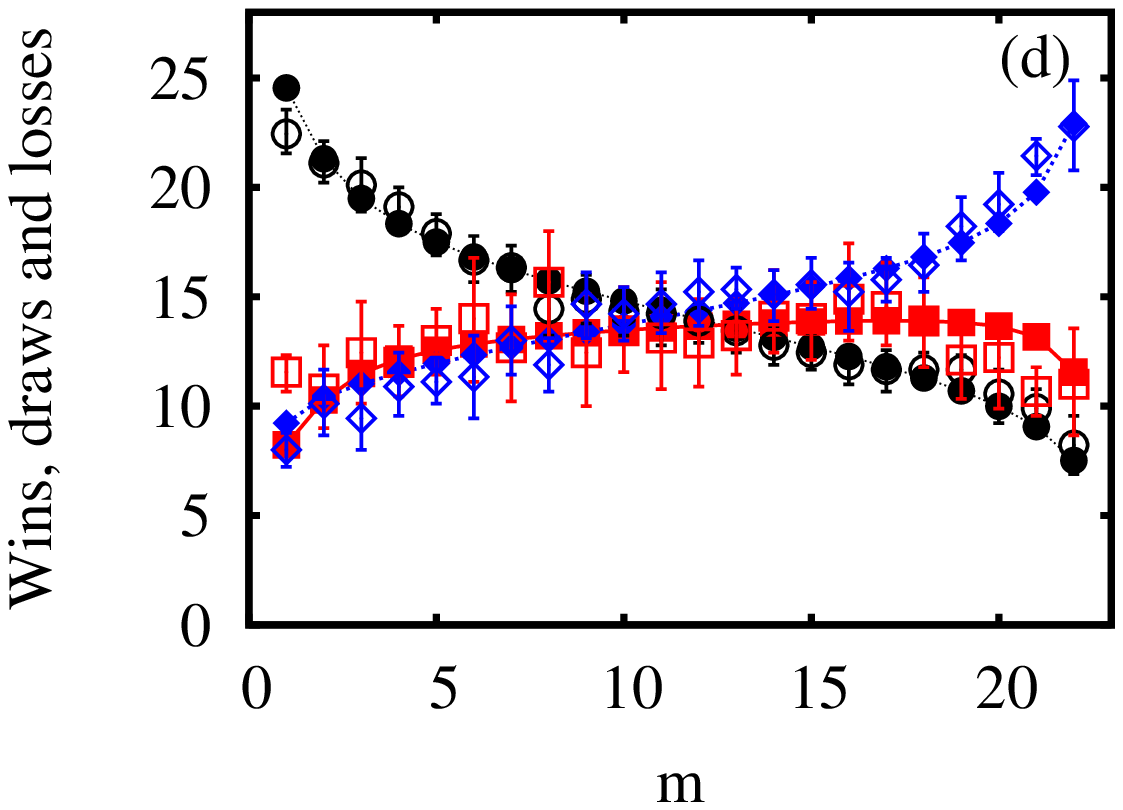}
	\caption{Number of wins (circles), draws (squares) and losses (diamonds) versus rank from observational data
(not filled markers) and non-identical team model (filled markers) for (a) German Bundesliga, (b) English  League, (c) Spanish League A and (d) Spanish League B.}
	\label{fig:vito-fit}
   \end{center}
\end{figure*}
Finally, we remark some aspects about the quality factor $Q(m)$ and mean number of points $\bar{s}(m)$.
Note that $\bar{s}(m)$ is the mean obtained from a very large number of tournaments, which does not coincide
with the mean of a tournament with a very large number of rounds. For instance, in a hypothetical tournament
with two (identical or not) teams playing only one game without draw we always have $\bar{s}(1)=3$ and $\bar{s}(2)=0$; in contrast, we obtain $\bar{s}(1)=1.5$ and $\bar{s}(2)=1.5$ when the number of matches goes to infinity for identical teams\cite{Ribeiro}.

Notice also that the result of a match depends on the quality of the involved teams, thus each $\bar{s}(m)$ would be function of all quality
factors. Therefore, each $Q(m)$ is a function of all $\bar{s}(m)$.
Unfortunately, is not an easy task to obtain a close form of $Q(m)$ in terms of $\bar{s}(m)$ due the non-linearity of the system of equations to be inverted. 
However, their shapes are in general similar, for example, Figure \ref{fig:quality}b shows this fact for the German Bunsdesliga when $r \rightarrow \infty$.
We also do not have a direct expression for the parameters $a$ and $b$ in eq. (\ref{eq:heuer}) in terms of $Q(m)$.

\section{Number of wins, draws and losses.}

Until now, we have analyzed only the teams' scores. Another perspective is to analyze the number 
of wins, draws and defeats. In order to do this, we evaluate the mean value of these quantities
over all the empirical data set and the emerging results from the non-identical team model. 
These results were obtained through simulations of an entire season using the
best fit parameters (Table \ref{tab:parameters}). In each simulation we counted (at the final round) 
the number of wins, draws and defeats of the rank $m$ and them we took an average over $10^5$ simulated seasons.
A comparison
between these two data can be found in Figure \ref{fig:vito-fit}. Although the model is based on
the scores, it gives a good agreement with the observational data. However, we observe that these variables
fluctuate more than the scores. This behavior is plausible, since the scores are constructed
as a linear combination of this three variables. 

We can see that the numbers of wins and defeats have a well defined hierarchical form. For instance,
in the case of the number of wins, it is greater for the first ranks and small for the last ones. However,
this hierarchical form is not clear when we look at the number of draws. The data behavior suggested
that the number of draws is almost constant over the rank positions.


\section{Comparison between tournament systems: an application}

When dealing with sports tournaments, one can ask about what kind of tournament system is better:
the all-play-all or the elimination tournaments? Here, we make an application of our model
and compare these tournament systems from a quantitative point of view.

In this context, we take the 16 best teams' quality factors which emerge from model after the adjustment.
Then, we use them to simulate an entire season $10^5$ times from both tournament systems.
Here, an entire season of the elimination tournaments consists of 4 rounds: the eight-finals,
quarter-finals, semi-finals and the final. In this kind of tournaments the loser of each match is 
immediately eliminated from the championship, for this reason, it is also referred
to as ``sudden death'' tournament.
\begin{figure*}[ht!]
   \begin{center}
	\includegraphics[scale=0.55]{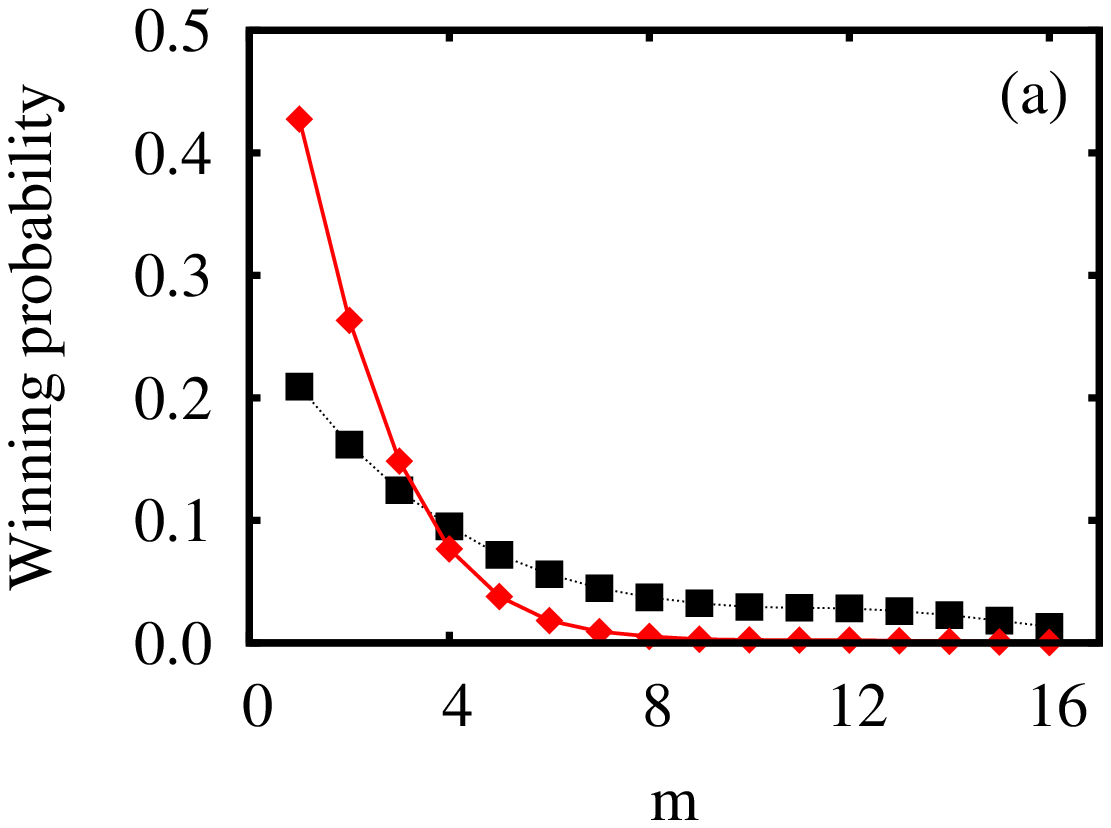}
	\includegraphics[scale=0.55]{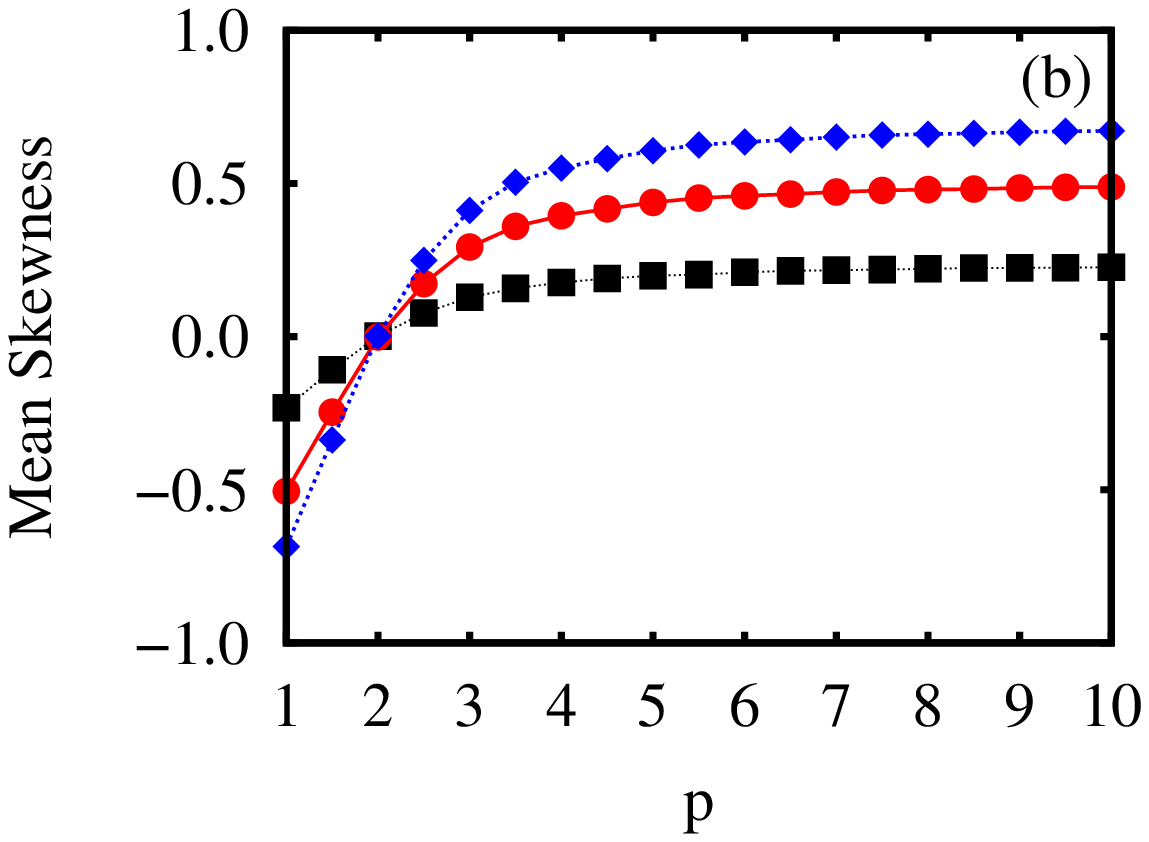}
	\caption{(a) Winning probability of a team with quality factor $Q(m)$ in an all-play-all (diamonds) tournament and  in an
		elimination (squares) tournament. We used the model parameters that emerge from the
		German Bundesliga data. (b) Mean skewness versus number
		of points for a win, $p$, for $\delta=0.2$ (squares), $\delta=0.4$ (circles) and 
		$\delta=0.4$ (diamonds).}
	\label{fig:prob}
   \end{center}
\end{figure*}
Employing this 
procedure and counting how many times a given team won the championship, we get the probability of winning 
$P(Q(m))$ in the two tournament systems. Figure \ref{fig:prob}a shows this probability as a function of the rank position by considering the German Bungesliga. Note that the teams with great quality factors
are considerably more likely to win in an all-play-all system than in an elimination one. On the other hand,
teams with small quality factor are more likely to win in an elimination tournament. This result indicates
that an elimination tournament has more randomness which enables less prepared teams to win. Unlike it,
the all-play-all system has more games and the randomness decreases which makes it less likely for a team
with small quality factor to win.

We also investigated the asymmetric tail characterized by positive skewness.
More specifically, we evaluated the mean value of the skewness over $10^5$ simulated seasons
for several numbers of points for a win, $p$ (as already pointed, $p=3$ corresponds to the current system).
In this simulation we use the identical team approach
and some values of $\delta$, as shown in Figure \ref{fig:prob}b. These results indicate that the asymmetric tail
is caused by the different score intervals between all possible results. In the most symmetric
way, i.e., no points for a defeat, $1$ point for a draw and $2$ for a win, the skewness is approximately
$0$ for all values of $\delta$. As $p$ increases, skewness also increases. However, for large values
of $p$ the mean skewness is approximately constant. When $p$ is large compared with $1$ 
(point for a draw), it dominates the score results and consequently the mean skewness saturates.
We can note that this plateau depends on the values of the parameter $\delta$, i.e.,
the plateau is larger for larger values of $\delta$.

\section{Summary}

In this work we investigated statistical aspects of soccer tournaments. The dynamics of these competitive systems was simulated by a simple probabilistic model, which retains relevant aspects of the leagues, such as the average rank score and the standard-deviations. Our results were compared with data from the German, the English and the Spanish soccer leagues and showed to be in good agreement with them.
Also, from these known data, the time evolution of the scores was studied as a random walk-like process.
These results indicated that the scores are not normally distributed, due to
the difference between the teams and the asymmetry of the scores system.
In addition, by using our model, we compared two tournament systems:
the all-play-all and the elimination tournaments. This comparison indicated that the eliminatory systems
have more randomness, which enables less prepared teams to win the tournament. In the all-play-all systems, the
randomness is smaller making the victory of the best teams more likely. 
In a more general context, due to a high degree of agreement between empirical data and the model,
we expect that the random walk-like model employed here may be useful to discuss other kind
of tournaments.

\begin{acknowledgement}
 The authors would like to thank CENAPAD-SP  (Centro Nacional de Processamento de
Alto Desempenho em S\~ao Paulo) for the computational support, and CNPq and CAPES
for partial financial support.
\end{acknowledgement}

\end{document}